\documentclass[twocolumn,showpacs,preprintnumbers,amsmath,amssymb]{revtex4}

\usepackage{graphicx}
\usepackage{dcolumn}
\usepackage{bm}
\usepackage{epstopdf}
\begin{document}

\preprint{APS/123-QED}

\title{Dewetting of Thin Viscoelastic Polymer
Films on Slippery Substrates}
\author{Thomas Vilmin}
\author{Elie Rapha{\"e}l}
\affiliation{
Laboratoire de Physique de la Mati\`ere Condens\'ee, FRE 2844 du CNRS, Coll\`ege de France, 11 Place Marcelin Berthelot, 75231 Paris Cedex 05, France
}

\date{\today}

\begin{abstract}
Dewetting of thin polystyrene films deposited onto silicone wafers at temperatures
close to the glass transition exhibits unusual dynamics and front morphologies. 
Here, we present a new theoretical approach of these phenomena taking into account 
both the viscoelastic properties of the film and the non-zero velocity of the film at the 
interface with the substrate (due to slippage). 
We then show how these two ingredients lead to : (a) A very asymmetric shape of the 
rim as the film dewetts,
(b) A decrease of the dewetting velocity with time like $t^{-\frac{1}{2}}$ for times shorter than the reptation time
(for larger times, the dewetting velocity reaches a constant value).
Very recent experiments by Damman, Baudelet and Reiter
[Phys. Rev. Lett. {\bf 91}, 216101 (2003)] present, 
however, a much faster decrease of the dewetting velocity. 
We then show how this striking result can be explained by the presence 
of residual stresses in the film.
\end{abstract}

\maketitle

Thin liquid films are of great scientific and technological importance, 
and display a variety of interesting dynamics phenomena \cite{PGGFBDQ, oron}.
In engineering, for instance, they serve to protect surfaces, and applications arise in
adhesives, magnetic disks and membranes. 
They have therefore been the focus of many experimental and theoretical 
studies \cite{revues}. When forced to cover a non-wettable substrate,
a thin liquid film is unstable and will dewet this substrate.
Four years ago, Reiter studied the dewetting of ultrathin ({\it i.e.} thinner than the coil size), 
almost glassy polystyrene (PS) films deposited onto silicon wafers coated with a
polydimethylsiloxane (PDMS) monolayer \cite{reiter2001}. 
He found that a highly asymmetric rim, 
with an extremely steep side towards the interior of the hole and a much slower decay 
on the rear side, builds up progressively. In order to explain these deviations from the behavior of simple
Newtonian liquids, theoretical models based on the shear-thinning properties of polymer films
\cite{Dalnoki1999} has been
proposed by Saulnier {\it et al.} \cite{saulnier2002} and Shenoy {\it et al.} \cite{shenoy2002}.
Very recently, Damman, Reiter and collaborators \cite{damman2003,reiter2003}
compared the opening of cylindrical holes with the retraction of a strait contact line.
These experiments revealed the existence of a highly asymmetric rim even in the case of the
edge geometry, and a strong decrease of the dewetting velocity as the rim builds up
(for both the radial and the edge geometries) \cite{Note}. One can show that none of these features 
can be explained by the existing theoretical models. 
In this letter, we therefore present a new theoretical approach that shows how the friction 
(due to slippage) of the liquid film onto the substrate, combined with the viscoelastic properties of 
the polymer film, can simply explain the above-mentioned experimental 
results of Damman and Reiter \cite{damman2003,reiter2003}.

\begin{figure}
\includegraphics[angle=90, clip=true, width = 7.4cm]{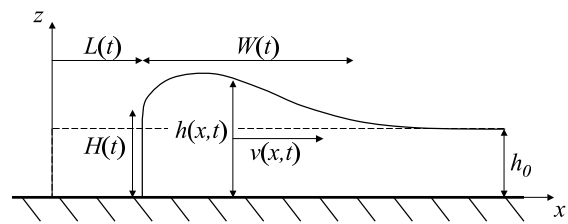}
\caption{\label{fig1} Film geometry : $h(r, t)$ is the profile of the film, $h_{0}$ is the initial height of the film, $H(t)$ is the height of the front, $L(t)$ is the dewetted distance, $W(t)$ is the width of the rim, and $v(x, t)$ is the velocity of the film.}
\end{figure}

Let us consider the dewetting of a straight contact line
and assume that the thickness of the liquid film 
is smaller than the hydrodynamic extrapolation length
$b = \eta / \zeta$ (where $\eta$ is the viscosity of the liquid, and $\zeta$ the friction coefficient 
of the liquid on the substrate) \cite{PGGFBDQ}. We can thus use a simple plug-flow description 
and characterize the velocity field in the film, $v(x, t)$, and the film profile,  $h(x, t)$,
by two functions independent of the $z$-coordinate (see Fig.~\ref{fig1}). 
The horizontal stress, $\sigma(x, t)$, is related to the strain rate, $\dot{\gamma} = {\partial v}/{\partial x}$, by a
constitutive equation (the form of which depends on the type of fluid under consideration; see below). 
Neglecting inertia, local mechanical equilibrium between the friction forces (per unit surface) onto the substrate, $\zeta \, v(x, t)$, and the bulk 
viscous forces
gives:
\begin{equation}
\zeta \,  v =\frac{\partial \left(h \sigma\right)}{\partial x}
\label{eq1}
\end{equation}
Assuming the fluid to be incompressible, volume conservation leads to:
\begin{equation}
\frac{\partial h}{\partial t} \, + \, vÊ\, \frac{\partial h}{\partial x} \; = \; 
-h \, \frac{\partial v}{\partial x}
\label{eq2}
\end{equation}
The last relation needed in order to solve the problem is the boundary condition at the edge of the film. 
The applied force (per unit of length) on the rim , $|S|$ (where $S$ is the spreading parameter \cite{PGGFBDQ} 
assumed to be negative), pushing the film away from the dry area, must be balanced by the
the viscous force:
\begin{equation}
|S| \; = \;  - H \, \sigma(x = L)
\label{eq3}
\end{equation}
where $H = H(t)$ is the front height, and $L = L(t)$ is the dewetted distance (see Fig.~\ref{fig1}).

\begin{figure}
\includegraphics[angle=-90, clip=true, width = 7.4cm]{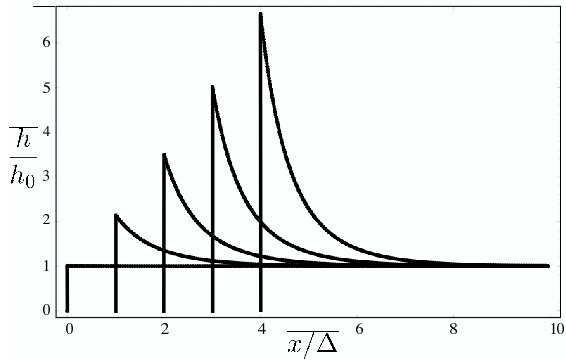}
\caption{\label{fig2} Numerical calculation of the shape of a  Newtonian film dewetting 
on a slippery substrate at different times. The different curves are separated by the same time interval.}
\end{figure}

For a Newtonian liquid, the above equations can easily be solved at short times.
Indeed, as long as  $h(x, t)$ remains of the same order as the initial thickness 
of the film, $h_{0}$, equation~(\ref{eq1}) - combined with the fact that for
a Newtonian fluid $\sigma \, = \, \eta \, \dot{\gamma}$ - leads to:
 \begin{equation}
\zeta \, v \, \simeq \, \eta \, h_{0} \, \frac{\partial^{2} v}{\partial x^{2}}
\label{eq2bis}
\end{equation}
The velocity field is then given by $v(x) \, = \, V_{0} \,  \exp{\left(-\frac{x-L}{\Delta}\right)}$,
where the distance $\Delta$ is given by $\sqrt{h_{0} \eta/\zeta} \, = \, \sqrt{h_{0} b}$, and the velocity $V_{0}$
by  $|S|/\sqrt{\zeta \eta h_{0}}$. At short times, a Newtonian liquid deposited on
a slippery substrate thus dewetts 
 with a constant velocity $V_{0}$. This result was already obtained by 
Brochard-Wyart {\it et al.} \cite{brochard97} using energetic arguments, but the present  
mechanical point of view gives us additional  informations about the film morphology. 
Indeed, Eqs.~(\ref{eq2}) and (\ref{eq3}) give for the front height~:
 $H = h_{0} +  (|S|/\eta)\,t$. Since the velocity field decreases exponentially
 as one moves away from the front, the film profile exhibits at short times
an asymmetric rim,
with an exponential decrease of the thickness over the characteristic length $\Delta$: 
\begin{equation}
h(x, t) \; = \; h_{0} \, +  \, \frac{|S|} {\eta}\,  t \exp{\left(-\frac{x - V_{0} t}{\Delta} \right)}
\label{eq4}
\end{equation}
This behavior is indeed the  observed by Reiter on AFM  images \cite{reiter2001}. 
We have also completed our analysis by numerically solving Eqs.~(\ref{eq1}), (\ref{eq2})
and (\ref{eq3}); as shown on Fig.~\ref{fig2}, these numerical solutions 
confirm well our analytical predictions. 
Our analysis also allows us to correct an assumption made 
by Brochard-Wyart {\it et al.} stating that the viscous dissipation should be
negligible compared with dissipation due to friction  \cite{brochard97}. Indeed,
a simple calculus based on the above results shows that the two dissipations are approximately equal.
Note that due to surface tension, the rim is in fact rounded over a distance 
 $\delta  \simeq H/\theta_{0}$ (where $\theta_{0}$ 
is the equilibrium contact angle). But the rim remains highly asymmetric if
 $\delta \ll \Delta$, that is to say as long as $H \ll \theta_{0} \Delta$, or, equivalently,
as long as $L \ll \theta_{0} b$. 
For $L \geq \theta_{0} b$, the friction of this cylindrical 
section on the substrate begins to be more important than the friction of the rest of the film, and simultaneously 
the Laplace pressure due to the curvature of the surface of the rim becomes stronger than the capillary pressure $|S|/H$. 
Thus, once  $L \geq \theta_{0} b$,  the "mature rim" regime described by Brochard-Wyart {\it et al.}\cite{brochard97}\cite{redon94} 
(see also Damman {\it{et al.}} \cite{damman2003}) begins, and the rim becomes round and symmetric (with a width
$W \sim \sqrt{h_{0} L}$ simply given by volume conservation). 
In this regime, the viscous dissipation is negligible compared with the dissipation due to friction,
and, consequently, the dewetting velocity is proportional to $t^{-\frac{1}{3}}$ \cite{brochard97}. 
Two important results arise from our analysis of the dewetting of a Newtonian fluid. 
Firstly, the friction of the film onto the substrate 
gives rise to an asymmetric rim, since it dumps the velocity field in the film over a length $\Delta$ (which 
depends on the liquid viscosity).
Secondly, the viscous dissipation
is approximately equal to interfacial dissipation due to friction during the formation of the rim,
while it is negligible in the "mature rim" regime. We can therefore anticipate that for 
a viscoelastic fluid, the rheologic properties of the fluid will have no significant consequences 
on the dewetting velocity in the "mature rim" regime, but will play a major role 
during the formation of the rim. 

\begin{figure}
\includegraphics[angle=-90, clip=true, width = 7.4cm]{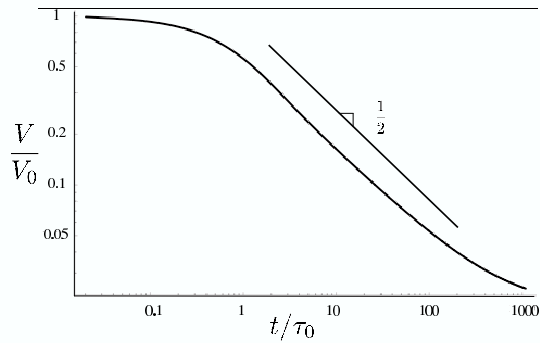}
\caption{\label{fig3} Numerical calculation of the reduced dewetting velocity $V/V_{0}$ versus 
the reduced time $t /\tau_{0}$ for a viscoelastic film with $\tau_{1} = 1000 \, \tau_{0}$. 
The straight line represents $(t/\tau_{0})^{-\frac{1}{2}}$.}
\end{figure}

Let us now consider in some details the dewetting of a viscoelastic film, 
assuming the following simplified constitutive equation \cite{bird}:
\begin{equation}
G \, \sigma \, + \, (\eta_{0} + \eta_{1}) \dot{\sigma} \; = \; G \, \eta_{1} \, \dot{\gamma} \; + \;  \eta_{0} \, \eta_{1} \, \ddot{\gamma}
\label{eq5}
\end{equation}
where $G$ is an elastic modulus (due to entanglements), $\eta_{0}$ is a short time viscosity
and $\eta_{1}$ is the usual melt viscosity
 ($\eta_{1} \gg \eta_{0}$). The time response of such a liquid can be divided
into three regimes: (1) At short times, $t < \tau_{0} = \eta_{0}/G$, the liquid behaves like a simple Newtonian liquid 
with weak viscosity $\eta_{0}$; (2) For $\tau_{0} < t < \tau_{1} =\eta_{1}/G$, where  
$\tau_{1}$ is the relaxation time of the liquid ({\it{i.e.}} the reptation time of the polymer chains)
the liquid behaves like an elastic solid of elastic modulus $G$. 
(3) At long times ($t > \tau_{1}$), the liquid behaves like a very viscous Newtonian liquid of viscosity $\eta_{1}$.
The above mentioned time response of the liquid has direct consequences 
on the dewetting process. 
For times shorter than $\tau_{0}$, the viscoelastic liquid dewets like a simple liquid, with a constant velocity 
$V_{0} = |S|/\sqrt{\zeta \eta_{0} h_{0}}$,  and with the formation 
of an asymmetric rim of width $\Delta_{0} = \sqrt{h_{0}\eta_{0}/\zeta}$.
At long times ($t > \tau_{1}$), the viscoelastic liquid also dewets like a simple liquid, with a constant velocity
$V_{1} = |S|/\sqrt{\zeta \eta_{1} h_{0}}$,  and with the formation 
of an asymmetric rim of width $\Delta_{1} = \sqrt{h_{0}\eta_{1}/\zeta} \gg \Delta_{0}$.
In between these two regimes, the viscoelastic behavior of the
fluid will thus lead to a significant  drop of the dewetting velocity (from $V_{0}$ to $V_{1}$). 
More precisely, in this intermediate time regime ($\tau_{0} < t < \tau_{1}$), the liquid behaves like an elastic solid and
the height of the front increases very slowly with time:  
$H \, \simeq \, h_{0} + |S| \left(1+t/\tau_{1}\right)/G \, \approx \, h_{0} + |S|/G$ \cite{RimHeight}. 
Volume conservation then imposes that the width of the rim, $W$, increases proportionally 
to the dewetted distance, $L$:
\begin{equation}
W \; \simeq \; \frac{G}{|S|} \, h_{0} \, L
\label{eq61}
\end{equation}
Let us now assume that
the bulk viscous dissipation is 
at most of the order of the dissipation due to friction. 
This assumption then allows us to determine the dynamic 
of the dewetting process from a simple energy balance (per unit of length) between the work done by the capillary force 
per unit of time and the dissipation due to friction:
\begin{equation}
|S| \, V \; \simeq \; \zeta \,  W \, V^{2}
\label{eq6}
\end{equation}
(where, as anywhere else in this letter, numerical prefactors of order unity are neglected).
The above equation, combined with Eq.~(\ref{eq61}), gives $V(t) \simeq V_{0} \sqrt{\tau_{0}/t}$ 
for times $t$ shorter than $\tau_{1}$. Note that this $t^{-\frac{1}{2}}$ behavior might have been
guessed directly from the fact that $V_{1}/V_{0} \, = \, \sqrt{\tau_{0}/\tau_{1}}$.
We have confirmed these results by numerically  solving the equations of motion
(see Fig.~\ref{fig3}). The good agreement between our analytical and numerical results indicates that our
assumption that the bulk viscous dissipation is smaller (or equal) to interfacial dissipation does hold for a viscoelastic liquid
described by Eq.~(\ref{eq5}).
For $t > \tau_{1}$, our model predicts a constant dewetting velocity $V_{1}$.
This prediction holds as long as  the height of the front $H$ is small compared 
to the width $\theta_{0} \Delta_{1}$. Thereafter, as in the case of a Newtonian liquid, 
the "mature rim" regime is reached
and the dewetting velocity then decreases like $t^{-\frac{1}{3}}$.
We have thus shown that for a viscoelastic fluid in the time interval 
$\tau_{0} < t < \tau_{1}$,
the width of the rim $W$ increases proportionally to the dewetted distance $L$,
and the dewetting velocity decreases like $t^{-\frac{1}{2}}$. The former prediction
is in good agreement with the experimental observations of Damman {\it et al.} \cite{damman2003}. 
While a $t^{-\frac{1}{2}}$ decrease of the dewetting velocity has indeed been occasionally observed
by Damman \cite{dammanperso}, in most cases the measured dewetting velocity decreases 
like $t^{-1}$ \cite{damman2003}, much faster than predicted.
In order to explain this striking result, 
let us assume that at the beginning of the dewetting process the polymer films are not at equilibrium, 
and display residual stresses due to the spin-coating fabrication process
and fast evaporation of the solvent, as recently emphasized by Reiter and De Gennes \cite{reiterpgg}. 
We shall now show how these residual stresses, assumed to be essentially horizontal 
and of initial amplitude  $\sigma_{0}$, cause a high initial dewetting velocity, followed by a strong slow down.

The various time regimes of the dewetting process, when partially driven by residual stresses, 
are similar to the ones already described when the process is only driven by capillary forces. 
At times shorter than $\tau_{0}$,
the dewetting velocity is equal to $V_{0}  \, + \, \sigma_{0} \Delta_{0}/ \eta_{0}$,
where the second term denotes the contribution of the residual stresses. In
this short times regime, the residual stresses have
no direct effects on the shape of the rim which thus keeps an exponential shape of characteristic width $\Delta_{0}$. 
For $t  > \tau_{0}$, the height of the front is given by
$H \, = \, h_{0} \, + \, h_{0} \sigma_{0}/G \, + \, |S| (1+t/\tau_{1})/G$ \cite{NewRimHeight},
and the width of the rim, $W$, is simply given by volume conservation: $W (H -  h_{0}) \, = \,  h_{0} L$.
Thus, as long as $t \ll \tau_{1}$, $H$ is approximately constant and $W$
increases proportionally to the dewetted distance. 
In order to obtain the dynamics of the dewetting process, the power (per unit of length) 
 $h_{0} \sigma_{0} \exp{(-\frac{t}{\tau_{1}})} \,V$ \cite{power} delivered by the residual stresses
should be added  to the {\it{l.h.s.}} of the energy balance Eq.~(\ref{eq6}). 
The dewetting velocity is then given by ($t > \tau_{0}$):
\begin{equation}
V \simeq  \frac{V_{1}}{\sqrt{2}} \frac{(1 + \epsilon + \frac{t}{\tau_{1}})(1 + \epsilon e^{-\frac{t}{\tau_{1}}})}{\sqrt{\frac{t}{\tau_{1}} + \frac{1}{2}(\frac{t}{\tau_{1}})^{2} + \epsilon(2 + \epsilon + \frac{t}{\tau_{1}})(1 - e^{-\frac{t}{\tau_{1}}})}}
\label{eq7}
\end{equation}
where $\epsilon = h_{0}\sigma_{0}/|S|$. 
Around $t = \tau_{0}$, the velocity decreases like $t^{-\frac{1}{2}}$, and thereafter decreases more sharply
as the residual stresses relax in the film. For large enough residual stresses ($\epsilon \geqslant 4$),
the dewetting velocity behaves like $t^{-1}$ around $t = 2\tau_{1}/3$.
Note that when the capillary forces are negligible ({\it{i.e.}} when $\epsilon \gg 1$), the residual stresses alone 
are able to induce the dewetting process and lead to a decrease of the dewetting velocity 
like $\exp{\left(-t/\tau_{1}\right)}$ (in the range $\tau_{1} \, < \, t < \, \tau_{1} \, \ln{\left(\epsilon(1+\epsilon)\right)}$.

\begin{figure}
\includegraphics[angle=-90, clip=true, width = 7.4cm]{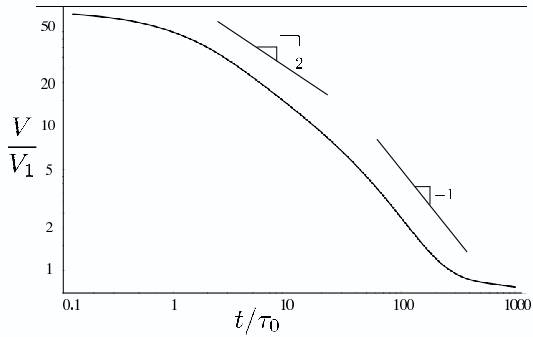}
\caption{\label{fig4} Numerical calculation of the reduced dewetting velocity, $V/V_{1}$,
versus reduced time, $t/\tau_{0}$,
for a viscoelastic film  ($\tau_{1} \, = \, 100 \, \tau_{0}$) with residual stresses 
($\sigma_{0} \, = \, G \, = \, 4 |S|/h_{0}$).
The straight lines represents $(t/\tau_{0})^{-1/2}$ and $(t/\tau_{0})^{-1}$ respectively.}
\end{figure}

The above analytical results are in good agreement with numerical solutions
(see Fig.~\ref{fig4}). Again, the simplified energy balance resulting from our assumption
that the bulk dissipation is smaller (or equal) to the interfacial dissipations gives very satisfying results.
Residual stresses are thus a very good candidate to explain the experimental observations 
of Reiter and Damman since both morphological observations (rim width $W$ proportional to the 
dewetted distance $L$), and dynamic measurements (variations with time of $L$ and $V$) are in 
very good agreement with the theoretical predictions. Additionally, it has been observed that
the dewetting velocity of a circular hole, while much lower 
than the dewetting velocity of a strait contact line at the beginning of the dewetting process
(due to radial deformations), systematically joins it after some time \cite{damman2003,reiter2003}.
The presence of residual stresses can simply explain this experimental observation. Indeed, as
described above, the dewetting velocity is mainly controlled by residual stresses (for $\epsilon > 1$)
which are evenly distributed throughout the film. Thus, when the size of a hole is large enough 
for the viscous dissipations due to radial deformations to be negligible compared with the friction onto
the substrate, the hole becomes equivalent to a strait line, and both velocities become of the same 
order, even-though the dewetted distances and the rim sizes are different in both cases. 

In conclusion, we have shown that the friction (due to slippage) of the liquid film 
onto the substrate can explain the building up of the asymmetric rim observed 
 by Reiter \cite{reiter2001} during the dewetting of thin PS films on a PDMS monolayer. 
We have also shown that the viscoelastic properties of the PS are of great importance as 
they lead to 
a decrease of the dewetting velocity with time proportional to $t^{-\frac{1}{2}}$ 
for times shorter than the reptation time of the polymer chains in the film. 
This decrease is made sharper by the presence of residual stresses.
A sharp decrease of the 
dewetting velocity  ($V \sim t^{-1}$) as observed by  Damman
{\it{et al.}} \cite{damman2003} could thus be seen as an evidence
of the presence of residual stresses in such viscoelastic films. 
Note that these residual stresses should also influence 
the initial stage of the opening of cylindrical holes (a situation where the dissipation due to radial deformations 
dominates over the friction).
The residual stresses might also play a role in the 
surface instabilities of the film and in the rate of hole formation \cite{reiterpgg}.
In this letter we did not talk about the shear thinning properties of PS films 
\cite{Dalnoki1999,saulnier2002}, but one can show - using an analysis 
similar to the one used in this letter for viscoelastic film - that a shear-thinning
behavior leads to a decrease of the dewetting velocity weaker
than $t^{-\frac{1}{2}}$. Hence, in the absence of residual stresses, 
shear thinning alone cannot explain the obeservations of Reiter,
Damman and colaborators,  even combined with viscoelastic properties.

\acknowledgments
We wish to acknowledge very interesting and fruitful discussions with G\"{u}nter Reiter and Pascal Damman.


\begin{thebibliography}{9}
\bibitem{RimHeight} This last formula is valid assuming $|S| < h_{0} G$. More generally, on has
$H \, = \, H_{0} \, + \, \frac{|S|}{G} \cdot \frac{t}{\tau_{1}}$ 
with $\frac{H_{0}}{h_{0}} \ln{\left(\frac{H_{0}}{h_{0}} \right)} \, = \, \frac{|S|}{h_{0} G}$.
\bibitem{NewRimHeight} 
This last formula is valid assuming $|S| \, + \, h_{0} \sigma_{0} \, < \, h_{0} G$.
A more general formula can be obtained along the lines of \cite{RimHeight}.
\bibitem{power}
The power delivered by the residual stresses is formally given by $\int{}{} h(x,t) \sigma(x,t) \dot{\gamma}(x,t) {\rm{d}}x$.
Using the fact that $\sigma(x,t)$ vanishes at the front ($x = L$), and reaches it unperturbed value
$\sigma_{0}\exp{(-\frac{t}{\tau_{1}})}$ far away from the rim ($x \gg W$), one can obtain the
simplified expression mentioned in the text.
\end{thebibliography}
\end{document}